# Electroluminescence emission from polariton states in GaAs-based semiconductor microcavities

A. A. Khalifa, A. P. D. Love, D. N. Krizhanovskii, M. S. Skolnick

*Department of Physics and Astronomy, University of Sheffield, Sheffield S3 7RH, United Kingdom*

J. S. Roberts

*Department of Electronic and Electrical Engineering, University of Sheffield, Sheffield S1 3JD, United Kingdom*

The authors report the observation of electroluminescence from GaAs-based semiconductor microcavities in the strong coupling regime. At low current densities the emission consists of two peaks, which exhibit anti-crossing behaviour as a function of detection angle and thus originate from polariton states. With increasing carrier injection we observe a progressive transition from strong to weak coupling due to screening of the exciton resonance by free carriers. The demonstration that polariton emission can be excited by electrical injection is encouraging for future development of polariton lasers.

During the last decade semiconductor microcavities have attracted great scientific interest[1]. In such structures strong exciton-photon coupling results in the formation of mixed exciton-photon quasiparticles, two-dimensional polaritons. Polaritons possess very small effective mass, ~ $10^{4-5}$ times less than that of bare excitons in quantum wells. As a result, it is possible to achieve high density macroscopically occupied polariton states at relatively high temperatures and small optical excitation densities. Recently, polariton condensation phenomena similar to that of Bose-Einstein condensation have been observed in CdTe[2] and GaAs[3] microcavities under condition of non-resonant optical pumping.

Such polariton condensation has significant potential for the development of low threshold coherent light sources based on semiconductor microcavities, so-called polariton lasers,[1] in which coherent emission arises from photon emission from the coherent, condensed state. Such stimulated emission is distinct from the stimulation in a photon laser since it is the polariton scattering process to the high density state which is stimulated, as opposed to the photon emission itself in a conventional laser. This separation of stimulation and emission leads to coherent emission without the requirement for population inversion, and to the prospect of very low threshold coherent sources. Such coherent emission has been reported under optical excitation and at temperatures less than ~30K in CdTe[2] and GaAs[3] systems. The recent observation of optically-excited stimulated emission at room temperature in GaN based microcavities in the strong coupling regime[4] is also highly encouraging.

For device applications, however, there is an important issue as to whether polariton emission and polariton lasing can be observed in semiconductor microcavities under conditions of electrical injection, where electrons and holes are injected separately from the opposite sides of the structure before relaxing and forming polariton states. This contrasts to polariton photoluminescence (PL), where the carriers are created optically, simultaneously at the same spatial location. Polariton dispersion in electroluminescence (EL) has been reported in microcavities containing an organic semiconductor as the active material[5]. However, up to now polariton emission and condensation in inorganic semiconductor microcavities has been observed only in the case of optical excitation. Recently, Bajoni et al reported EL from a single InGaAs QW inserted in the intrinsic region of a *p-i-n* photodiode[6]. A narrow emission line, corresponding to exciton emission has been observed, confirmed by photoluminescence excitation spectra which revealed a narrow excitonic resonance at low injection currents[6].

In this letter we report the observation of EL from GaAs/AlGaAs based microcavities in the strong coupling regime embedded in a *p-i-n* junction. At low current density we observe emission consisting of two peaks, which correspond to the lower and upper polariton branches. Angular resolved EL measurements reveal a strong dependence of the energy of the peaks on the detection angle, which show marked anti-crossing. In addition, the dispersion of the lower polariton is observed to have a point of inflection at an angle of ~15 degrees. Such observations are characteristic of the occurrence of exciton-photon strong coupling in the system. With increasing voltage (and, thus, current) applied to the sample the strong coupling becomes less pronounced: both upper and lower polariton peaks exhibit broadening and the energy splitting between them decreases. At high currents the splitting collapses due to screening of the exciton resonance. The EL then consists of a single broad line at the energy of the bare photon mode and corresponds to the emission of an electron-hole plasma in the weak coupling regime.

The sample studied was a $3\lambda/2$ GaAs cavity, incorporating two sets of three 100Å $In_{0.06}Ga_{0.94}As$ quantum wells separated by 100Å GaAs barriers, positioned at the antinodes of the optical field. The top (bottom) mirror consisted of 17 (20) repeats of $\lambda/4$ layers of AlAs and $Al_{0.18}Ga_{0.82}As$.[7] Doping was introduced into the mirror regions to form a *p-i-n* junction, so that an electric field could be applied across the intrinsic cavity region. A gold contact was deposited on the back side of the sample. Circular mesas with diameter $d$~300 μm were processed using optical lithography and wet chemical etching. Ti–Au ring contacts were defined on top of the mesas. The sample is maintained at low temperature (10 K) using a cold finger cryostat. Several mesas were investigated with Rabi splitting $\Omega \sim 6$ meV and near zero detuning between exciton and cavity modes. A HeNe laser was focused to ~70 µm on the sample for comparative PL measurements. The angular dependence of EL and PL was studied using a fiber-coupled goniometer. The emission from a ~70 µm region on the mesa was detected with a 0.85 m double spectrometer and a CCD camera.

Fig.1 (a) shows a typical current-voltage I-V dependence of the *p-i-n* diode in forward bias ($V_B$). Onset of significant current flow occurs at 2-2.5V. This is somewhat greater than the expected turn-on voltage of 1.5-2V in GaAs-based structures, probably due to voltage-drop across the Bragg mirrors which were not optimized to achieve minimum series resistance.[8]

Fig.1 (b) shows emission spectra recorded at detection angles of $\theta \sim 0^0$ with respect to the normal to the microcavity plane under conditions of electrical (top) injection of free carriers with current of 5.3 mA, and optical excitation (bottom) with a HeNe laser of 1mW power. The excitation power densities are estimated to be around 0.2 µW/µm$^2$ in both the EL[9] and PL cases taking into account the size of the electrical mesa (d~300 µm) and that of the laser excitation spot (~70 µm). In both cases the spectra are observed to be very similar with similar emission intensities (within a factor of 2): each spectrum consists of two peaks at energies ~1.4539 eV and ~1.46 eV, corresponding as we show below to the emission of lower (LP) and upper (UP) branch polaritons, respectively. The weak background emission between the two peaks probably arises from localized exciton states uncoupled to the cavity mode.[7] The very similar EL and PL emission intensities in fig.1 (b) indicate that the efficiencies of the relaxation processes for carriers injected either optically or electrically are very similar for our structures.

To prove that the observed peaks originate from polariton emission we investigated the EL and PL as a function of detection angle θ. Fig.2 (a) shows EL spectra recorded as a function of θ in the range from $-5^0$ to $+40^0$. Both the LP and UP peaks exhibit a pronounced angular dependence. The LP and UP dispersions obtained from EL (open) are plotted in fig.2 (b). The LP dispersion has a point of inflection at ~15 degrees, continues to increase with angle and then tends towards constant energy at higher detection angles ~$40^0$. We also observe pronounced anti-crossing between the LP and the UP emission (Fig.2(a), (b)). The strong angular dependence of both the LP and UP peaks, the observation of the point of inflection and constant energy at high angle of the LP peak provide direct evidence for the occurrence of exciton-photon strong coupling in the system. In addition, we investigated the polariton dispersion in the case of optical excitation only (Fig. 2 (b), solid). The dispersion obtained from PL is found to be identical to that obtained from EL. Such an observation is additional evidence for polariton formation under electrical pumping, since it is well known from previous studies that similar microcavity structures[7] show polariton emission in the strong coupling regime under optical excitation. Furthermore very good fits to the observed dispersion curves are obtained (solid lines in fig.2 (b)) assuming strong coupling between quantum well excitons and the photonic mode.[10] From the fits, we deduce the Rabi splitting and the detuning between uncoupled photonic and exciton modes at k=0 to be ~6 meV and –1.3 meV, respectively. We note that if there were not bound excitonic states in the system under electrical injection, the system would be in the weak coupling regime. In this case a quadratic angular dispersion would be observed with a single peak without a point of inflection, corresponding to emission from the bare photonic mode.

The current densities of ~3 A/cm$^2$ at which we observe polariton EL are similar to the current densities reported in Ref[6], at which EL of excitons in quantum wells has been observed. If we assume that all the injected carriers relax and then recombine in the quantum well, using an expression for the relationship between concentration of free carriers and excitons in thermal equilibrium[11] we estimate the exciton densities $n_X$ in our structures to be around ~1-2x10$^9$ cm$^{-2}$ per quantum well at currents ~3 mA. This is about one order of magnitude less than the exciton densities (~5x10$^{10}$ cm$^{-2}$) at which the Mott transition between excitons and electron-hole pairs is expected.[12]

Finally, we performed EL studies as a function of applied voltage (current). Fig.3 (a) shows EL spectra recorded for currents in the range from 0.6 mA to 25 mA at zero detection angle. Fig 3 (b) shows that the emission intensity of the UP and LP peaks (peak heights) to a very good approximation is a linear function of current at currents < 10 mA, showing the minimal influence of non-radiative effects. At higher voltages (currents) both UP and LP lines exhibit broadening, whilst the splitting between them decreases from 6 meV at 0.66 mA to 4.7 meV at 13 mA. There is also a slight red shift of the LP peaks by ~1 meV, probably arising from sample heating of the sample with increasing current. From the shift we estimate that the temperature increases up to ~30 K. At ~ 25 mA both the LP and UP peaks have disappeared and the emission is replaced by a single broad line with a maximum at the energy of the bare photonic mode 1.456 eV at $\theta=0^0$ (Fig.2 b). Such behavior is indicative of screening of the excitonic resonance and a transition from the strong to weak coupling regimes due to increase of the concentration of free carriers in the system at higher currents. We note that the collapse of strong coupling occurs at currents of ~25 mA, corresponding to exciton densities about 1-2x10$^{10}$ cm$^{-2}$ per quantum well. This is consistent with the excitonic densities at which the Mott transition is expected to occur (~5x10$^{10}$ cm$^{-2}$).[12]

Although we do not observe stimulated emission in the present structures before the loss of strong coupling, the prospects to achieve polariton lasing under conditions of electrical injection remain good. A polariton laser may be realized in GaAs structures with a high number of quantum wells, which allows reduction of the concentration of excitons per quantum well at the threshold polariton density for stimulated emission.[3] However, recently Bajoni[13] et al have cast some doubt on this approach for GaAs-based structures, in contrast to ref 3: in the regime of stimulated emission under optical excitation they observed a parabolic dispersion from GaAs microcavities with a large number of quantum wells (12-15), indicating that the stimulated emission occurs in the weak coupling regime. By contrast, there is no doubt that polariton condensation does occur in CdTe microcavities with their larger exciton binding energies and hence greater resistance to screening[2]: such systems could be used for the creation of an electrically pumped polariton laser. Also GaN microcavities, where stimulated emission at 300K has been observed recently[4] in cavities in the strong coupling regime under non-resonant optical excitation are very promising, although at the expense of a more complex materials technology.[14]

In conclusion, we have observed polariton electroluminescence from GaAs-based semiconductor microcavities in the strong coupling regime. The strong coupling regime is verified by the observation of angular dispersion for both lower and upper polariton with characteristic anti-crossing behavior. Our results are encouraging for the development of low-threshold polariton lasers in appropriately designed structures and materials combinations. Finally, we note that after submission of our manuscript we became aware of the other research reporting EL from GaAs microcavities in the strong coupling regime[15].

Acknowledgements: The work was supported by EU projects Clermont 2 RTN-CT-2003-503677 and Stimscat 517769, and EPSRC grants GR/S09838/01 and GR/S76076/01. D. N. Krizhanovskii is an EPSRC Advanced Fellow (EP/E051448). We thank R. But
té for suggestions which led to the commencement of this work.

**Figure captions:**

Fig.1: (a) Current-voltage dependence of the microcavity *p-i-n* diode at 10 K. (b) Emission spectra recorded at $\theta \sim 0^0$ for electrical (top) injection with current of 5.3 mA and optical excitation (bottom) with HeNe laser of 1mW power.

Fig.2: (a) EL spectra as a function of detection angle $\theta$ in the range from $-5^0$ to $+40^0$. (b) Dispersion of LP and UP peaks observed in EL (open) and PL (solid). The full lines are a fit to the observed dispersions using a two mode model of coupled exciton and photon oscillators. The dispersions of the uncoupled photon and exciton modes deduced from the fit to the polariton dispersions are given by the dotted lines.

Fig.3: (a) EL spectra as function of current. (b) The intensity of LP and UP peaks versus current.

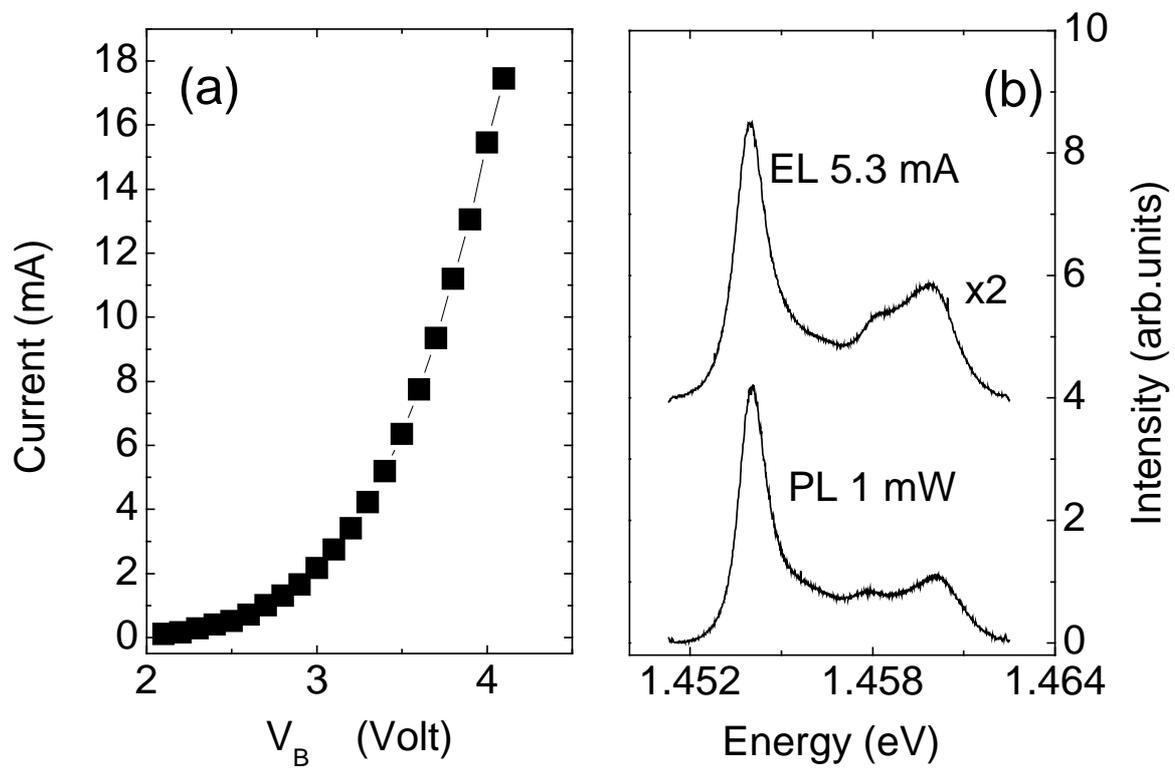

Fig.1

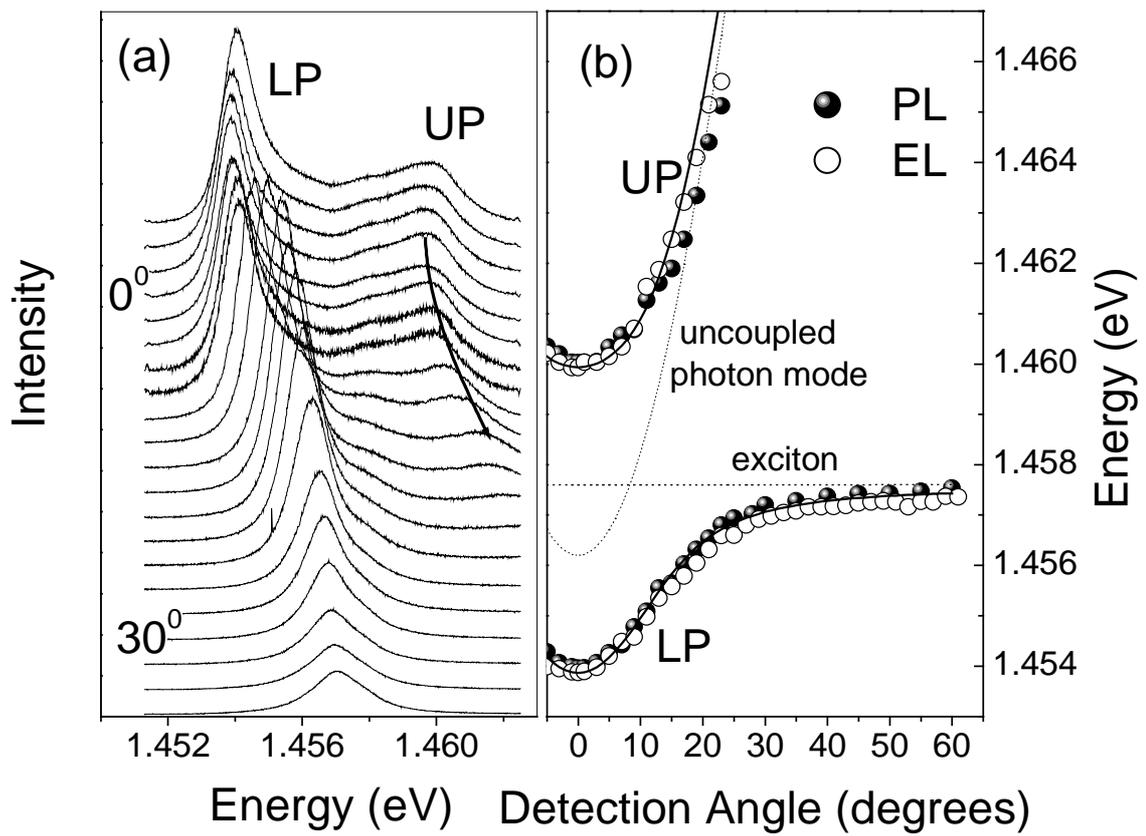

Fig.2

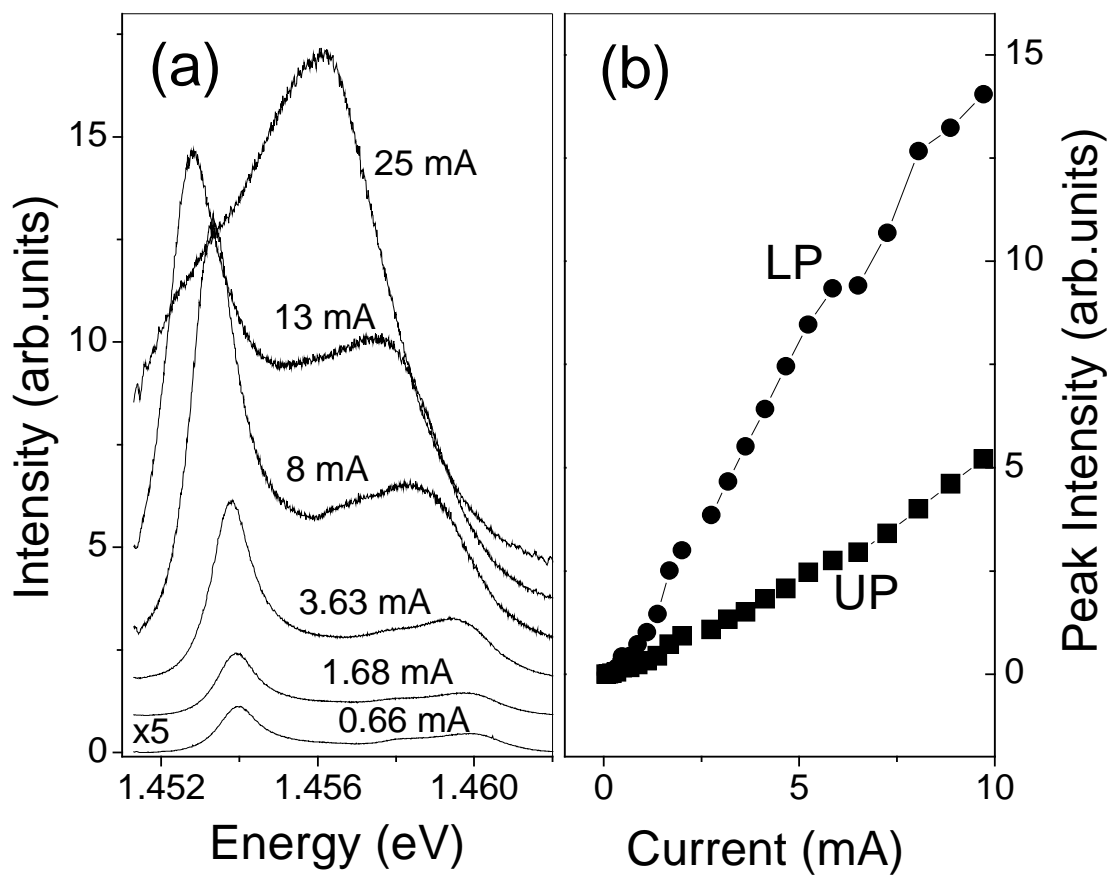

Fig.3